\documentclass{elsart}
\usepackage{hyperref}
\usepackage{amssymb}
\usepackage[dvips]{graphicx}% Include figure files
\usepackage{dcolumn}% Align table columns on decimal point
\usepackage{bm}% bold math
\usepackage{amsmath}

\begin{document}

\begin{frontmatter}


\title{Detecting the quantum zero-point motion of vortices in the cuprate superconductors}

\author[Harvard,Frankfurt]{Lorenz Bartosch},
\author[UCSB]{Leon Balents}, and
\author[Harvard]{Subir Sachdev}

\address[Harvard]{Department of Physics, Harvard University, Cambridge MA
02138}

\address[Frankfurt]{Institut f\"ur Theoretische Physik, Universit\"at
Frankfurt, 60054 Frankfurt, Germany}

\address[UCSB]{Department of Physics, University of California, Santa
Barbara, CA 93106-4030}

%\author{Subir Sachdev}
%\address{Department of Physics, Yale University,\\ P.O. Box 208120,
%New Haven, CT 06520-8120, USA} \ead{subir.sachdev@yale.edu}
%\ead[url]{http://pantheon.yale.edu/\~\/subir}

\begin{abstract}
We explore the experimental implications of a recent theory of the
quantum dynamics of vortices in two-dimensional superfluids
proximate to Mott insulators. The theory predicts modulations in the
local density of states in the regions over which the vortices
execute their quantum zero point motion. We use the spatial extent
of such modulations in scanning tunnelling microscopy measurements
(Hoffman {\em et al\/}, Science {\bf 295}, 466 (2002)) on the vortex
lattice of Bi$_2$Sr$_2$CaCu$_2$O$_{8+\delta}$ to estimate the
inertial mass of a point vortex. We discuss other, more direct,
experimental signatures of the vortex dynamics.
\end{abstract}

%\begin{keyword} Mott insulator \sep bond order \sep
%fractionalization \sep Ising gauge theory

%\PACS
%\end{keyword}
\end{frontmatter}

\section{Introduction}

It is now widely accepted that superconductivity in the cuprates is
described, as in the standard Bardeen-Cooper-Schrieffer (BCS)
theory, by the condensation of charge $-2e$ Cooper pairs of
electrons. However, it has also been apparent that vortices in the
superconducting state are not particularly well described by BCS
theory. While elementary vortices do carry the BCS flux quantum of
$hc/2e$, the local electronic density of states in the vortex core,
as measured by scanning tunnelling microscopy (STM) experiments, has
not been explained naturally in the BCS framework. Central to our
considerations here are the remarkable STM measurements of Hoffman
{\em et al.} \cite{hoffman} (see also Refs.~\cite{fischer,pan}) who
observed modulations in the local density of states (LDOS) with a
period of approximately 4 lattice spacings in the vicinity of each
vortex core of a vortex lattice in
Bi$_2$Sr$_2$CaCu$_2$O$_{8+\delta}$.

This paper shall present some of the physical implications of a
recent theory of two-dimensional superfluids in the vicinity of a
quantum phase transition to a Mott insulator \cite{bbbss1,bbbss2}
(see also Ref.~\cite{zlatko}). By `Mott insulator' we mean here an
incompressible state which is pinned to the underlying crystal
lattice, with an energy gap to charged excitations. In the Mott
insulator, the average number of electrons per unit cell of the
crystal lattice, $n_{MI}$, must be a rational number. If the Mott
insulator is not `fractionalized' and if $n_{MI}$ is not an even
integer, then the Mott insulator must also spontaneously break the
space group symmetry of the crystal lattice so that the unit cell
of the Mott insulator has an even integer number of electrons.
There is evidence that the hole-doped cuprates are proximate to a
Mott insulator with $n_{MI} = 7/8$ \cite{jtran}, and such an
assumption will form the basis of our analysis of the STM
experiments on Bi$_2$Sr$_2$CaCu$_2$O$_{8+\delta}$. The electron
number density in the superfluid state, $n_S$, need not equal
$n_{MI}$ and will be assumed to take arbitrary real values, but
not too far from $n_{MI}$.

\begin{figure}
\centering
\includegraphics[width=4.5in]{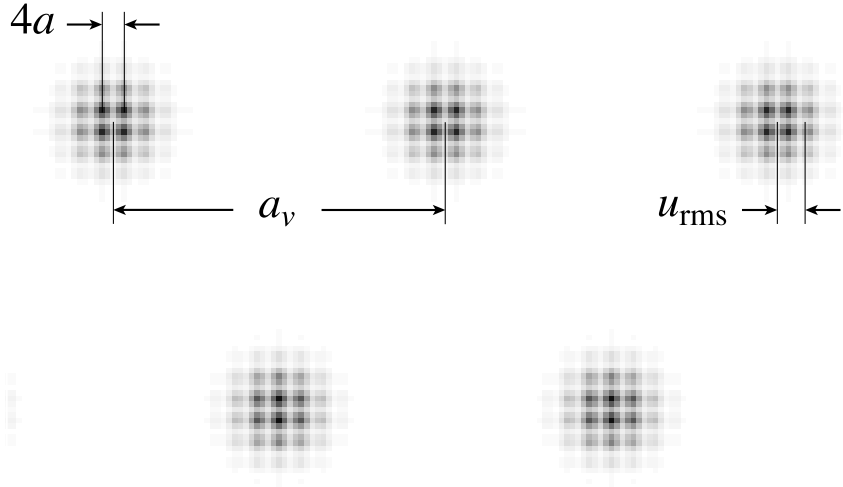}
\caption{Schematic of the modulations in the LDOS of a vortex
lattice. The short distance modulations in each vortex halo are
determined by the orientation of the vortex in flavor space, as
discussed in Ref.~\cite{bbbss1}. The envelope of these modulations
is $|\Psi ({\bf r}_j) |^2$ where $\Psi$ is the wavefunction of the
vortices, and its characteristics are computed in the present
paper.} \label{figfluxlattice}
\end{figure}

A key ingredient in our analysis will be the result that the
superfluid carries a subtle quantum order, which is distinct from
Landau-Ginzburg order of a Cooper pair condensate. In two
dimensions, vortices are point-like excitations, and are therefore
bona fide quasiparticle excitations of the superfluid. The quantum
order is reflected in the wavefunction needed to describe the
motion of the vortex quasiparticle. For $n_{MI}$ not an even
integer, the low energy vortices appear in multiple degenerate
flavors, and the space group symmetry of the underlying lattice is
realized in a projective unitary representation that acts on this
flavor space. Whenever a vortex is pinned (either individually due
to impurities, or collectively in a vortex lattice), the space
group symmetry is locally broken, and hence the vortex necessarily
chooses a preferred orientation in its flavor space. As shown in
Ref.~\cite{bbbss1}, this implies the presence of modulations in
the LDOS in the spatial region over which the vortex executes its
quantum zero point motion \cite{vbs}. The short-distance structure
and period of the modulations is determined by that of the Mott
insulator at density $n_{MI}$, while its long-distance envelope is
a measure of the amplitude of the vortex wavefunction (see
Fig.~\ref{figfluxlattice}).
Consequently, the size of the region where the modulations are
present is determined by the inertial mass of the vortex. Here we
will show how these ideas can be made quantitatively precise, and
use current experiments to obtain an estimate of the vortex mass,
$m_v$. There have been a number of theoretical discussions of
$m_v$ using BCS theory
\cite{volovik,kopnin,blatter,duan,tvr,Coffey98,han},
and they lead to the order of magnitude estimate $m_v \sim m_e
(k_F \xi)^2$, where $m_e$ is the electron mass, $k_F$ is the Fermi
wavevector, and $\xi$ is the BCS coherence length.

\section{Vortex equations of motion}
\label{sec:eom}

We begin with a very simple, minimal model computation of the vortex
dynamics, in which retardation, dissipation, and inter-layer Coulomb
interactions will be neglected. This serves the purpose of exposing
the basic physics. The following section will present a much more
complete derivation, in which these effects will be re-instated, and
the connection to the field theory analysis of Refs.~\cite{bbbss1}
will also be made explicit.

Consider a system of point vortices moving in a plane at positions
${\bf r}_j$, where $j$ is a label identifying the vortices. We do
not explicitly identify the orientation of each vortex in flavor
space, because we are interested here only in the long-distance
envelope of the LDOS modulations; the flavor orientation does not
affect the interactions between well-separated vortices, and so
plays no role in determining the wavefunction of the vortex lattice.
In a Galilean-invariant superfluid, the vortices move under the
influence of the Magnus force
\begin{equation}
m_v \frac{d^2 {\bf r}_j}{d t^2} = \frac{h n_S}{2 a^2} \left( {\bf
v}_s ({\bf r}_j ) - \frac{d {\bf r}_j}{dt} \right) \times \hat{\bf
z}, \label{e1}
\end{equation}
where $t$ is time, $h=2\pi \hbar$ is Planck's constant, ${\bf v}_s
({\bf r})$ is the superfluid velocity at the position ${\bf r}$,
and $n_S/a^2$ is the electron number density per unit area ($a^2$
is the area of a unit cell of the underlying lattice). One point
of view is that the force in Eq.~(\ref{e1}) is that obtained from
classical fluid mechanics after imposing the quantization of
circulation of a vortex. However, Refs.~\cite{haldane} emphasized
the robust topological nature of the Magnus force and its
connection to Berry phases, and noted that it applied not only to
superfluids of bosons, but quite generally to superconductors of
paired electrons. Here, we need the modification of Eq.~(\ref{e1})
by the periodic crystal potential and the proximate Mott
insulator. This was implicit in the results of Ref.~\cite{bbbss1},
and we present it in more physical terms. It is useful to first
rewrite Eq.~(\ref{e1}) as
\begin{equation}
m_v \frac{d^2 {\bf r}_j}{d t^2} = {\bf F}_E (j) + {\bf F}_B (j),
\label{e2}
\end{equation}
where ${\bf F}_E$ is the first term proportional to ${\bf v}_s$
and ${\bf F}_B$ is the second term. Our notation here is
suggestive of a dual formulation of the theory in which the
vortices appear as `charges', and these forces are identified as
the `electrical' and `magnetic' components. In the Galilean
invariant superfluid, the values of ${\bf F}_E$ and ${\bf F}_B$
are tied to each other by a Galilean transformation. However, with
a periodic crystal potential, this constraint no longer applies,
and their values renormalize differently as we now discuss.

The influence of the crystal potential on ${\bf F}_E$ is simple,
and replaces the number density of electrons, $n_S$, by the
superfluid density. Determining ${\bf v}_s ({\bf r}_j )$ as a sum
of contributions from the other vortices, we obtain
\cite{background}
\begin{equation}
{\bf F}_E (j) = 2 \pi \rho_s \sum_{k (\neq j)} \frac{{\bf r}_j - {\bf
r}_k}{|{\bf r}_j - {\bf r}_k |^2} \;, \label{e3}
\end{equation}
where $\rho_s$ is the superfluid stiffness (in units of energy).
It is related to the London penetration depth, $\lambda$, by
\begin{equation}
\rho_s = \frac{\hbar^2 c^2 d}{16 \pi e^2 \lambda^2},
\end{equation}
where $d$ is the interlayer spacing.

The modification of ${\bf F}_B$ is more subtle. This term states
that the vortices are `charges' moving in a `magnetic' field with
$n_S/2$ `flux' quanta per unit cell of the periodic crystal
potential. In other words, the vortex wavefunction is obtained by
diagonalizing the Hofstadter Hamiltonian which describes motion of
a charged particle in the presence of a magnetic field and a
periodic potential. As argued in Ref.~\cite{bbbss1}, it is useful
to examine this motion in terms of the deviation from the rational
`flux' $n_{MI}/2=p/q$ ($p$, $q$ are relatively prime integers)
associated with the proximate Mott insulator. The low energy
states of the rational flux Hofstadter Hamiltonian have a $q$-fold
degeneracy, and this constitutes the vortex flavor space noted
earlier \cite{sflux}. However, these vortex states describe
particle motion in zero `magnetic' field, and only the deficit
$(n_{S} - n_{MI})/2$ acts as a `magnetic flux'. This result is
contained in the action in Eq.~(2.46) of Ref.~\cite{bbbss1}
(see also Eq.~(\ref{s0p}) below), which
shows that the dual gauge flux fluctuates about an average flux
determined by $(n_{S} - n_{MI})$. The action in Ref.~\cite{bbbss1}
has a `relativistic' form appropriate to a system with equal
numbers of vortices and anti-vortices. Here, we are interested in
a system of vortices induced by an applied magnetic field, and can
neglect anti-vortices; so we should work with the corresponding
`non-relativistic' version of Eq.~(2.46) of Ref.~\cite{bbbss1}. In
its first-quantized version, this `non-relativistic' action for
the vortices leads to the `Lorentz' force in Eq.~(\ref{e2}) given
by
\begin{equation}
{\bf F}_B (j) =  -\frac{h (n_S  -n_{MI})}{2 a^2}
\frac{d {\bf r}_j}{dt} \times \hat{\bf z}, \label{e4}
\end{equation}
If the density of the superfluid equals the commensurate density of
the Mott insulator, then ${\bf F}_B = 0$; however, ${\bf F}_E$ remains
non-zero because we can still have $\rho_s \neq 0$ in the superfluid.
These distinct behaviors of ${\bf F}_{E,B}$ constitute a key
difference from Galilean-invariant superfluids. In experimental
studies of vortex motion in superconductors \cite{ong}, a force of the
form of Eq.~(\ref{e4}) is usually quoted in terms of a `Hall drag'
co-efficient per unit length of the vortex line, $\alpha$;
Eq.~(\ref{e4}) implies
\begin{equation}
\alpha = - \frac{h (n_S - n_{MI})}{2a^2 d}. \label{e5}
\end{equation}
Thus the periodic potential has significantly reduced the magnitude of
$\alpha$ from the value nominally expected \cite{kopkra} by
subtracting out the density of the Mott insulator. A smaller than
expected $|\alpha|$ is indeed observed in the cuprates \cite{ong}.  It
is worth emphasizing that ${\bf F}_B$ (but not ${\bf F}_E$) is an {\sl
  intrinsic} property of a single vortex.  Moreover, we expect that,
taken together, the relation Eq.~(\ref{e5}) and the flavor degeneracy
$q$ are robust ``universal'' measures of the quantum order of a clean
superconductor, independent of details of the band structure, etc.

\section{Derivation from field theory}
\label{sec:ft}

We will now rederive the results of the previous section from a more
sophisticated perspective. We will use a field theoretic approach to
derive an effective action for the vortices, a limiting case of
which will be equivalent to the equations of motion already
presented. The effective action will include retardation effects,
and can be easily extended to include inter-layer interactions and
dissipation.

Our starting point is a model of ordinary bosons on the square lattice
interacting via the long-range Coulomb interaction. Following
Ref.~\cite{bbbss1} we will briefly review a duality mapping of this model
into a field theory %\cite{bbbss1}
for vortices in a superfluid of bosons
which is in the vicinity of a transition to a Mott insulator.
%We will pay special attention to the long-range Coulomb interaction to
%be considered here.
%Also, we will relate coupling constants in our dual theory to coupling
%constants in the original model.
The density of bosons per unit cell of the
underlying lattice is $\rho_B = n_B/a^2$, while the density of bosons in the Mott
insulator is $\rho_{MI} = n_{MI}^{(B)}/a^2 = (p/q)/a^2$; here
%$p$ and $q$ are relatively prime integers, and
$a^2$ is the unit cell area of the underlying
lattice. Closely related field theories apply to models of electrons
on the square lattice appropriate to the cuprate superconductors
\cite{bbbss2,bs}, with the boson density replaced by the
corresponding density of Cooper pairs; the needed extensions do not
modify any of the results presented below.
It should be noted that since two electrons pair to form one
Cooper pair the average number of electrons in the Mott insulating state
is $n_{MI} = 2 n_{MI}^{(B)}$ (and the average number of electrons in the
superfluid phase is $n_s = 2 n_B$).

In zero applied magnetic field, the Hamiltonian of our system is given by
\begin{equation}
\mathcal{H} = - \rho_s \sum_{i\alpha} \cos \left(\Delta_{\alpha} \hat{\phi}_i
% - 2 \pi g_{i\alpha}
\right)
% \nonumber \\
%&~&~~~~~~~~
+ \frac{e^{\ast \, 2}}{2} \sum_{i\neq j} \frac{(\hat n_i - n_B)(\hat n_j -n_B)}{|{\bf r}_i - {\bf r}_j|} \;,
\label{eq:Hamiltonian}
\end{equation}
where $\rho_s$ is the superfluid stiffness and $-e^{\ast}$ ($= -2 e$) is the charge of a boson (Cooper pair).
The bosons are represented by conjugate rotor and number operators
$\hat{\phi}_i$ and $\hat{n}_i$ which live on the sites $i$ of the square lattice (with position vector ${\bf r}_i$) and satisfy the
commutation relations
\begin{equation}
[\hat{\phi}_i, \hat{n}_j] = i \delta_{ij} \;.
\end{equation}
We subtract the average boson density $n_B$ from the number operators $\hat n_i$ to
account for global charge neutrality of the system.
Finally, we have introduced the discrete lattice derivative
$\Delta_\alpha \hat{\phi}_i = \hat{\phi}_{i+\alpha} - \hat{\phi}_i$ along one of the
two spatial directions $\alpha = x$ or $y$.

\subsection{Dual lattice representation}
\label{sec:dual}

Let us now briefly review the duality analysis of the above model with
special emphasis on the long-range Coulomb interaction.
%Also, we will relate coupling constants in our dual theory to coupling
%constants in the original model.
Following
Ref.~\cite{bbbss1} we represent the
partition function of $\mathcal{H}$ as a Feynman path integral
by inserting complete sets of eigenstates to the number operators $\hat n_i$
at times separated by the imaginary time slice $\Delta \tau$.
While the Coulomb interaction term is diagonal in this basis, the
hopping term in $\mathcal{H}$ can be easily evaluated by making use of the
Villain representation
\begin{equation}
  \exp\left(\rho_s \Delta\tau \cos \left(\Delta_{\alpha} \hat{\phi}_{i} \right)\right)
  \rightarrow \sum_{\{J_{i\alpha}\}}
  \exp \left( - \frac{J_{i\alpha}^2}{2 \rho_s \Delta \tau} + i J_{i
      \alpha} \Delta_{\alpha} \hat{\phi}_{i} \right) \;. \label{eq:Villain}
\end{equation}
Here, we have set $\hbar =1$ and have dropped an unimportant normalization
constant which we will also do in the following.
The $J_{i \alpha}$ are integer
variables residing on the links of the direct lattice,
representing the current of the bosons.

Extending the lattice index
$i$ to spacetime and introducing the integer-valued boson current in spacetime,
$J_{i\mu} \equiv (n_i, J_{ix}, J_{iy})$, the partition function can be written as
\begin{equation}
\mathcal{Z} = {\sum_{\{J_{i\mu}\}}}^{\prime} \exp \left( - \frac{1}{2\Delta\tau\rho_s}
\sum_{i\alpha} J_{i \alpha}^2 - \frac{\Delta\tau\, e^{\ast\, 2}}{2}
\sum_{i\neq j} \frac{(J_{i0} - n_B)(J_{j0} -n_B)}{|{\bf r}_i - {\bf r}_j|}
\right) \;,
\end{equation}
where the prime on the sum over the $J_{i\mu}$ restricts this sum to configurations
satisfying the continuity equation
\begin{equation}
  \label{eq:continuity}
  \Delta_{\mu} J_{i\mu} = 0 \;.
\end{equation}
This constraint can explicitly be solved by writing
\begin{equation}
J_{i\mu} = \epsilon_{\mu\nu\lambda} \Delta_\nu A_{\bar i \lambda} \;,
\end{equation}
where $A_{\bar i \mu}$ is an integer-valued gauge field on the links of the dual
lattice with lattice sites $\bar i$. We can now
promote $A_{\bar i \mu}$ from an integer-valued field to a real field by
the Poisson summation method. We then soften the integer
constraint with a vortex fugacity $y_v$ and make the
gauge invariance of the dual theory explicit by
%introducing the phase field $\vartheta_{\bar i}$
by replacing
$2 \pi A_{{\bar i} \mu }$ by $2 \pi A_{{\bar i} \mu} -
\Delta_{\mu} \vartheta_{\bar i}$.
The operator $e^{i \vartheta_{\bar i}}$ is
then the creation operator for a vortex in the boson phase variable
$\phi_i$. We now arrive at the dual partition function
\begin{align}
 \mathcal{Z}_d = \,& \prod_{\bar i} \int d A_{{\bar i} \mu} \int d \vartheta_{\bar i}\,
\exp \Biggl(y_v \sum \cos\left(2\pi A_{\bar i \mu} - \Delta_{\mu} \vartheta_{\bar i}\right) \nonumber \\
&\qquad {} - \frac{1}{2\Delta\tau \rho_s} \sum \left(
\epsilon_{\alpha\nu\lambda} \Delta_{\nu} A_{{\bar i} \lambda} \right)^2
\nonumber \\
&\qquad {} - \frac{\Delta\tau\, e^{\ast\, 2}}{2}
\sum \frac{(\epsilon_{0\nu\lambda} \Delta_{\nu} A_{\bar i \lambda} - n_B)
(\epsilon_{0\nu'\lambda'} \Delta_{\nu'} A_{\bar j \lambda'} -n_B)}{|{\bf r}_i - {\bf r}_j|}
 \Biggr) \;. \label{zdv}
\end{align}
As a last step we can replace the hard-core vortex field $e^{i \vartheta_{\bar i}}$
by the ``soft-spin'' vortex field $\psi_{\bar i}$, resulting in
\begin{align}
  \mathcal{Z}_d = \,& \prod_{\bar i} \int d A_{{\bar i} \mu} \int d \psi_{\bar i}\,
\exp \Biggl(\frac{y_v}{2}
\sum \left[ \psi_{\bar i + \mu}^{\ast} e^{2 \pi i A_{\bar i \mu}}
\psi_{\bar i} +
\mbox{c.c.} \right] -\sum \left[ s |\psi_{\bar i} |^2 + \frac{u}{2} |\psi_{\bar i}|^4
\right]
\nonumber \\
&\qquad {} - \frac{1}{2\Delta\tau \rho_s} \sum \left(
\epsilon_{\alpha\nu\lambda} \Delta_{\nu} A_{{\bar i} \lambda} \right)^2
\nonumber \\
&\qquad {} - \frac{\Delta\tau\, e^{\ast\, 2}}{2}
\sum \frac{(\epsilon_{0\nu\lambda} \Delta_{\nu} A_{\bar i \lambda} - n_B)
(\epsilon_{0\nu'\lambda'} \Delta_{\nu'} A_{\bar j \lambda'} -n_B)}{|{\bf r}_i - {\bf r}_j|}
 \Biggr) \;. \label{zdvs}
\end{align}
The first two terms in the exponent describe the action of the vortex fields
$\psi_{\bar i}$ which are minimally coupled to the gauge field $A_{\bar i \mu}$.
While the system is in a superfluid
phase for $s \gg 0$ it is in a Mott insulating phase for $s \ll 0$.

At boson filling $n_{MI}^{(B)}=p/q$ the gauge field $A_{\bar i \mu}$
in the action in Eq.~(\ref{zdvs}) fluctuates around the saddle point
$\bar A_{\bar i \mu}$ with
$\epsilon_{\mu\nu\lambda} \Delta_{\nu} A_{\bar i \lambda} = n_{MI}^{(B)} \delta_{\mu,\tau}$. It is therefore customary to substitute the gauge field $A_{\bar i \mu}$ by
\begin{align}
  A_{\bar i \tau} \to \,& \bar A_{\bar i \tau} +\frac{\Delta\tau}{2\pi} A_{\bar i \tau} \;, \\
  A_{\bar i \alpha} \to \,& \bar A_{\bar i \alpha} +\frac{a}{2\pi} A_{\bar i \alpha} \;.
\end{align}
Here we have already rescaled the deviations of the gauge field from the saddle
point such that later on we can easily take the continuum limit.
A careful analysis of the symmetry properties of the above dual vortex theory
(see Ref.~\cite{bbbss1}) shows that the vortex fields transform under a
projective symmetry group whose representation is at least $q$-fold degenerate.
It was also argued in Ref.~\cite{bbbss1} that while $q$ cannot be chosen too large
the boson density in the superfluid phase $n_B$ can take any value not too far
away from the boson density in the Mott insulating phase, $n_{MI}^{(B)}$.

In zero applied magnetic field, and at a generic boson density
$\rho_B$, the field theory for such a superfluid is then given by
%appears in Eqn.~(2.46)
%of Ref.~\cite{bbbss1}, and by focusing on the quadratic order terms about the
%saddle point of Eq.~(\ref{zdvs}) we also obtain from the above (with $\hbar = 1$)
\begin{align}
&\mathcal{S}_\varphi = \int d^2 r d \tau \left( \sum_{\ell =
0}^{q-1} \left[ |(\partial_\mu - i A_{\mu} ) \varphi_\ell |^2 +
m_v^2 |\varphi_\ell |^2 \right] +\frac{1}{8 \pi^2 \rho_s}
\left(\nabla A_\tau -
\partial_\tau {\bf A} \right)^2 \right) \nonumber \\
& \qquad {} + \frac{e^{\ast 2}}{8\pi^2}
  \int d^2 r \int d^2 r' \int d \tau \nonumber \\
& {} \times \frac{\left( \hat{\bf z} \cdot (\nabla
\times {\bf A}({\bf r},\tau)) - 2 \pi (\rho_B - \rho_{MI}) \right)
\left( \hat{\bf z} \cdot (\nabla
\times {\bf A}({\bf r}',\tau)) - 2 \pi (\rho_B - \rho_{MI}) \right)}
{|{\bf r} - {\bf r}'|} \nonumber \\
& \qquad {} + \ldots
\label{s0p}
\end{align}
This equation is a modified version of Eq.~(2.46) in Ref.~\cite{bbbss1}
with the short-range interaction between bosons replaced by the
long-range Coulomb interaction.
$\varphi_\ell$ is a vortex field operator which is the sum of a
vortex annihilation and an anti-vortex creation operator, and $\ell$
is the vortex flavor index. As discussed above, as long as the
vortices are well separated, the flavor index $\ell$ plays no role
in determining the zero-point motion of the vortices, and hence the
envelope of the modulations illustrated in Fig~\ref{figfluxlattice};
we will therefore drop the flavor index in the subsequent
discussion. Recall that the index $\mu$ runs over the spacetime co-ordinates $\tau$,
$x$, $y$
(while the index $\alpha$ runs only over the spatial co-ordinates $x$, $y$).
We have rescaled the $\tau$ co-ordinate so that the
`relativistic velocity' appearing in the first term  is unity.

The vortices in $\mathcal{S}_\varphi$ are coupled to a non-compact
U(1) gauge field $A_\mu = (A_\tau, {\bf A})$. The central property
of boson-vortex duality is that the `magnetic' flux in this gauge
field, $\hat{\bf z} \cdot (\nabla \times {\bf A})/(2 \pi)$ is a
measure of the boson density. However, notice from the last term in
$\mathcal{S}_\varphi$ with co-efficient $e^{\ast 2}$
%($U$ is the strength of
%the short-range repulsive interaction between the bosons)
that the
action is minimized by an average gauge flux (or boson density) of
$(\rho_B - \rho_{MI})$, the deviation in the density from that of the
Mott insulator, and {\em not\/} at the total boson density $\rho_B$,
as one would expect from usual considerations of the Magnus
force on continuum superfluids. The origin of this shift in the
average flux is explained in detail in Ref.~\cite{bbbss1}; briefly
stated, the combination of the periodic potential and `magnetic'
flux acting on the vortices has the effect of transmuting the flux
associated with the density of the Mott insulator into the $q$
vortex flavors. Only the deficit $\rho_B - \rho_{MI}$ then acts as a
`magnetic' field on the vortices.

The vortices also experience an `electric' field, whose fluctuations
are controlled in the action $\mathcal{S}_\varphi$ by the boson
superfluid density $\rho_s$.

The vortex component of the action $\mathcal{S}_\varphi$ has a
`relativistic' form and so describes both vortices and anti-vortices
with vanishing net mean vorticity. We are interested here in the
case of a vortex lattice induced by an applied (real, not dual)
magnetic field. In the dual language, this magnetic field appears as
a static background `charge' density which interacts via the
`electric' force with the `charged' vortices and anti-vortices.
Finiteness of energy requires that this background charge density
induces a neutralizing density of vortex `charges', which, in the
classic Abrikosov theory, form a vortex lattice (in the dual
language this lattice is a Wigner crystal of charges). We will
neglect anti-vortices from now on, and focus only on the dynamics of
these vortices induced by the applied field. For the action
$\mathcal{S}_\varphi$ this restriction means that we should work
with the `non-relativistic' limit. The formal procedure for taking
this limit was discussed in Section IV.B of Ref.~\cite{bbbss1}, and
leads to an action for a non-relativistic field $\Psi$, which is a
vortex annihilation operator (anti-vortices have been eliminated
from the spectrum). As shown earlier, the action for $\Psi$ takes
the form
\begin{align}
& \mathcal{S}_\Psi = \int d^2 r d \tau \left( \Psi^\ast
(\partial_\tau-i A_\tau) \Psi  + \frac{1}{2 m_v} |(\nabla -i {\bf A}
) \Psi |^2 + \frac{1}{8 \pi^2 \rho_s} \left(\nabla A_\tau -
\partial_\tau {\bf A} \right)^2 \right)  \nonumber \\
& \qquad {} + \frac{e^{\ast 2}}{8\pi^2}
  \int d^2 r \int d^2 r' \int d \tau \nonumber \\
& {} \times \frac{\left( \hat{\bf z} \cdot (\nabla
\times {\bf A}({\bf r},\tau)) - 2 \pi (\rho_B - \rho_{MI}) \right)
\left( \hat{\bf z} \cdot (\nabla
\times {\bf A}({\bf r}',\tau)) - 2 \pi (\rho_B - \rho_{MI}) \right)}
{|{\bf r} - {\bf r}'|} \nonumber \\
& \qquad {} + \dots
\label{s2p}
\end{align}
We now transform from this second quantized form of the vortex
action to a first quantized form with vortices at spatial positions
${\bf r}_j (\tau)$ where, as before, $j$ is a vortex label. In this
form the action is
\begin{align}
& \mathcal{S}_R = \int d\tau \sum \frac{m_v}{2} \left( \frac{d{\bf
r}_j}{d \tau} \right)^2 + \int d^2 r d \tau \Biggl( i A_\tau \rho +
i{\bf A} \cdot {\bf J} +  \frac{1}{8 \pi^2 \rho_s} \left(\nabla
A_\tau -
\partial_\tau {\bf A} \right)^2  \nonumber \\
& \qquad {} + \frac{e^{\ast 2}}{8\pi^2}
  \int d^2 r \int d^2 r' \int d \tau \nonumber \\
& {} \times \frac{\left( \hat{\bf z} \cdot (\nabla
\times {\bf A}({\bf r},\tau)) - 2 \pi (\rho_B - \rho_{MI}) \right)
\left( \hat{\bf z} \cdot (\nabla
\times {\bf A}({\bf r}',\tau)) - 2 \pi (\rho_B - \rho_{MI}) \right)}
{|{\bf r} - {\bf r}'|} \nonumber \\
& \qquad {} + \ldots
\label{s1p}
\end{align}
where $\rho$ and ${\bf J}$ are vortex density and currents respectively:
\begin{align}
\rho ({\bf r}, \tau) = & \, \sum \delta({\bf r} - {\bf r}_j (\tau)) \;, \nonumber \\
{\bf J } ({\bf r}, \tau) = & \, \sum \frac{d {\bf r}_j}{d \tau}
\delta({\bf r} - {\bf r}_j (\tau)) \;.
\end{align}

Now it is useful to shift the vector potential $A_{\alpha}$ to absorb the
mean background flux
\begin{equation}
{\bf A} \rightarrow \frac{\hat{\bf z} \times {\bf r}}{2} B + {\bf
A} \;,
\end{equation}
where
\begin{equation}
B \equiv 2 \pi (\rho_B - \rho_{MI}) \;.
\end{equation}
The fluctuations of the flux about this average value are controlled
by the long-range Coulomb
interactions. We assume that the vortices are located
near the positions of a regular vortex lattice with equilibrium
positions ${\bf R}_j$, and make displacements ${\bf u}_j$ from these
positions such that ${\bf r}_j = {\bf R}_j + {\bf u}_j$.
Adopting the Coulomb gauge, $\nabla \cdot {\bf A} = 0$,
the resulting action for the vortices is $\mathcal{S}_u =
\mathcal{S}_1 + \mathcal{S}_2 + \mathcal{S}_3$ where
\begin{align}
\mathcal{S}_1 = &\, \int d \tau \sum_j \left( \frac{m_v}{2}
\left(\frac{d{\bf u}_j}{d \tau} \right)^2 + i \frac{B}{2} \hat{\bf
z} \cdot \left( {\bf u}_j \times \frac{d{\bf u}_{j}}{d
\tau}\right) \right) \;, \nonumber \\
\mathcal{S}_2 = & \, \int \frac{d^2 q d \omega}{8 \pi^3} \Biggl(
\frac{1}{8 \pi^2 \rho_s} \left[ q^2 |A_\tau (q, \omega) |^2 +
\omega^2 | {\bf A} (q, \omega) |^2 \right] + \frac{e^{\ast 2} }{4\pi
q} |{\bf q} \times {\bf A} ({\bf q}, \omega)|^2
 \Biggr) \;, \nonumber \\
\mathcal{S}_3 = & \, \int d^2 r d \tau \left( i A_\tau \rho + i{\bf J}
\cdot {\bf A} \right) \;.
\end{align}
It is interesting to note that all couplings in this action are
known from experiments, apart from the vortex mass $m_v$.

Now we integrate out the $A_\tau$ and ${\bf A}$, and expand the
resulting action carefully to second order in the ${\bf u}$. (We
also use the component notation $u_{\alpha}$, where the index ${\alpha}$ extends
over the $x$ and $y$ components.)
 This
directly yields the result
\begin{equation}
\mathcal{S} = \frac{1}{2} \sum_{\alpha,\beta} \int \frac{d \omega}{2 \pi}
\int_{1{\rm BZ}} \frac{d^2 q}{4 \pi^2} u_\alpha (-{\bf q}, -\omega)
D_{\alpha\beta} ({\bf q}, \omega) u_\beta ({\bf q}, \omega) \;, \label{u1}
\end{equation}
where the momentum integral is over the first Brillouin zone of the
vortex lattice,
\begin{equation}
{\bf u} ({\bf q}, \omega) = \int d \tau \sum_j {\bf
u}_j e^{-i {\bf q} \cdot {\bf R}_j
 + i \omega \tau} \;,
\end{equation}
and the
dynamical matrix is
\begin{eqnarray}
D_{\alpha\beta} ({\bf q}, \omega) &=& A_0 m_v \omega^2 \delta_{\alpha\beta} + A_0
\omega B \epsilon_{\alpha\beta} - \sum_{{\bf G} \neq 0} \frac{4 \pi^2 \rho_s
G_\alpha G_\beta}{|{\bf G}|^2} \nonumber \\ &+& \sum_{{\bf G}}  \frac{ 4
\pi^2 \rho_s(q_\alpha + G_\alpha) (q_\beta + G_\beta)}{ |{\bf q}+{\bf G}|^2 + \omega^2
|{\bf q}+{\bf G}|/(2 \pi \rho_s e^{\ast 2})}
\nonumber \\
  &+& \delta_{\alpha\beta} \sum_{{\bf G}}  \frac{ 4 \pi^2
\rho_s\omega^2}{ \omega^2 + 2 \pi \rho_s e^{\ast 2} |{\bf q}+{\bf
G}|} \;, \label{u2}
\end{eqnarray}
where $A_0$ is the area of a unit cell of the vortex lattice, and
${\bf G}$ extends over all the reciprocal lattice vectors of the
vortex lattice of points ${\bf R}_j$.

It is now not difficult to show (see Appendix \ref{appendix:Ewald})
that, after dropping retardation
effects, the action in Eqs. (\ref{u1}, \ref{u2}) is equivalent to
the harmonic equations of motion that would be obtained for the
vortex lattice from Eqs.~(\ref{e2}-\ref{e4}). Instantaneous
interactions are obtained by taking the $e^\ast \rightarrow \infty$
limit of Eq.~(\ref{u2}). Clearly, the present formalism allows us to
include these without much additional effort.

So far, the action is free from dissipation effects associated with
the Bardeen-Stephen viscous drag. We will consider these in
Section~\ref{sec:viscous} below. For now we note that these can be
included in the above action simply by the transformation
\begin{equation}
D_{\alpha\beta} ({\bf q}, \omega) \rightarrow D_{\alpha\beta} ({\bf q}, \omega)  +
\delta_{\alpha\beta}\, \eta\, d \, |\omega|. \label{Deta}
\end{equation}
As we will see in Eq. (\ref{e9}), $\eta$ is the viscous drag
co-efficient, and $d$ is the spacing between the layers.

The present formalism also allows us to consider the coupling
between different two-dimensional layers in the cuprate system, and
this will be examined in the following subsection.

\subsection{Interlayer Coulomb interactions}
\label{sec:coulomb}

Even in the absence of any Josephson or magnetic couplings between
the layers, it is clear that we at least have to account for the
interlayer Coulomb interactions because the vortex spacing is much
larger than the layer spacing $d$.

We create a copy of all fields in all layers, labelled by the layer
index $n$. In particular, we now have gauge fields $A_\mu^{(n)}$.
The Coulomb couplings between the layers modify $\mathcal{S}_2$ to
\begin{align}
\mathcal{S}_2 = & \, \int \frac{d \omega}{2 \pi} \int_{1{\rm BZ}}
\frac{d^2 q}{4 \pi^2} \Biggl( \frac{1}{8 \pi^2 \rho_s}
\sum_{n}\left[ q^2 |A_\tau^{(n)} ({\bf q}, \omega) |^2 +
\omega^2 | {\bf A}^{(n)} ({\bf q}, \omega) |^2 \right] \nonumber \\
&~~~~+ \frac{e^{\ast 2}}{4 \pi q } \sum_{n,n'} e^{-|n-n'| qd}
({\bf q} \times {\bf A}^{(n)} (-{\bf q},-\omega)) \cdot ({\bf q}
\times {\bf A}^{(n')} ({\bf q}, \omega)) \Biggr) \;.
\end{align}
The interlayer interaction comes from the Fourier transform of
$1/\sqrt{r^2 + (n-n')^2 d^2}$. Now we perform a Fourier transform of
the layer index, into a momentum perpendicular to the layer,
$p_\perp$, leading to the field $A_\mu ({\bf q}, \omega, p_\perp)$.
In terms of this field
\begin{align}
\mathcal{S}_2 = &\, \int_{-\pi}^{\pi} \frac{ d p_\perp}{2 \pi} \int
\frac{d \omega}{2 \pi} \int_{1{\rm BZ}} \frac{d^2 q}{4 \pi^2}
\Biggl( \frac{1}{8 \pi^2 \rho_s} \sum_{n}\left[ q^2 |A_\tau ({\bf
q}, \omega, p_\perp) |^2 +
\omega^2 | {\bf A} ({\bf q}, \omega, p_\perp) |^2 \right] \nonumber \\
& {} + \frac{e^{\ast 2}}{4\pi q} \frac{1 - e^{-2 qd}}{1+ e^{-2 q d} - 2
e^{-q d} \cos p_\perp}({\bf q} \times {\bf A} (-{\bf q},-\omega,
-p_\perp)) \cdot ({\bf q} \times {\bf A} ({\bf q}, \omega, p_\perp))
\Biggr) \;.
\end{align}
Because we always have $qd \ll 1$, we can simplify this to
\begin{align}
\mathcal{S}_2 = &\, \int_{-\pi}^{\pi} \frac{ d p_\perp}{2 \pi} \int
\frac{d \omega}{2 \pi} \int_{1{\rm BZ}} \frac{d^2 q}{4 \pi^2}
\Biggl( \frac{1}{8 \pi^2 \rho_s} \sum_{n}\left[ q^2 |A_\tau ({\bf
q}, \omega, p_\perp) |^2 +
\omega^2 | {\bf A} ({\bf q}, \omega, p_\perp) |^2 \right] \nonumber \\
& {} + \frac{e^{\ast 2}d}{4\pi(1 - \cos p_\perp)}({\bf q} \times {\bf
A} (-{\bf q},-\omega, -p_\perp)) \cdot ({\bf q} \times {\bf A} ({\bf
q}, \omega, p_\perp)) \Biggr) \;. \label{s2}
\end{align}
Now, as before, we integrate out the $A_\mu$ and obtain the
effective action for the ${\bf u}$ in a single layer. This has the
form as in Eq.~(\ref{u1}), but with the dynamical matrix in
Eq.~(\ref{u2}) replaced by
\begin{eqnarray}
D_{\alpha\beta} ({\bf q}, \omega) &=& A_0 m_v \omega^2 \delta_{\alpha\beta}
+ A_0 \omega B \epsilon_{\alpha\beta} - \sum_{G \neq 0} \frac{4 \pi^2  \rho_s G_i G_j}{|{\bf G}|^2} \nonumber \\
&+& \sum_{{\bf G}} \int_{-\pi}^{\pi} \frac{ d p_\perp}{2 \pi} \frac{
4 \pi^2  \rho_s(q_\alpha + G_\alpha) (q_\beta + G_\beta)}{ |{\bf q}+{\bf G}|^2 +
\omega^2 (1- \cos p_\perp )/(2 \pi \rho_s e^{\ast 2} d)} \nonumber
\\
&+& \delta_{\alpha\beta} \sum_{{\bf G}} \int_{-\pi}^{\pi} \frac{ d p_\perp}{2
\pi} \frac{ 4 \pi^2 \rho_s\omega^2 (1 - \cos p_\perp )}{ \omega^2 (1
- \cos p_\perp ) + 2 \pi \rho_s e^{\ast 2} |{\bf q}+{\bf G}|^2 d } \;.
\label{dabd}
\end{eqnarray}

\section{Vortex lattice normal modes}
\label{sec:vlat}

We begin by presenting the numerical solution of the minimal model
presented in Section~\ref{sec:eom}. This is equivalent to solving
the dynamical matrix in Eq.~(\ref{u2}) in the limit $e^\ast
\rightarrow \infty$. The influence of all the additional effects
considered in Section~\ref{sec:ft} will be described in the next
section.

\begin{figure}[t]
\centering
\includegraphics[width=4.5in]{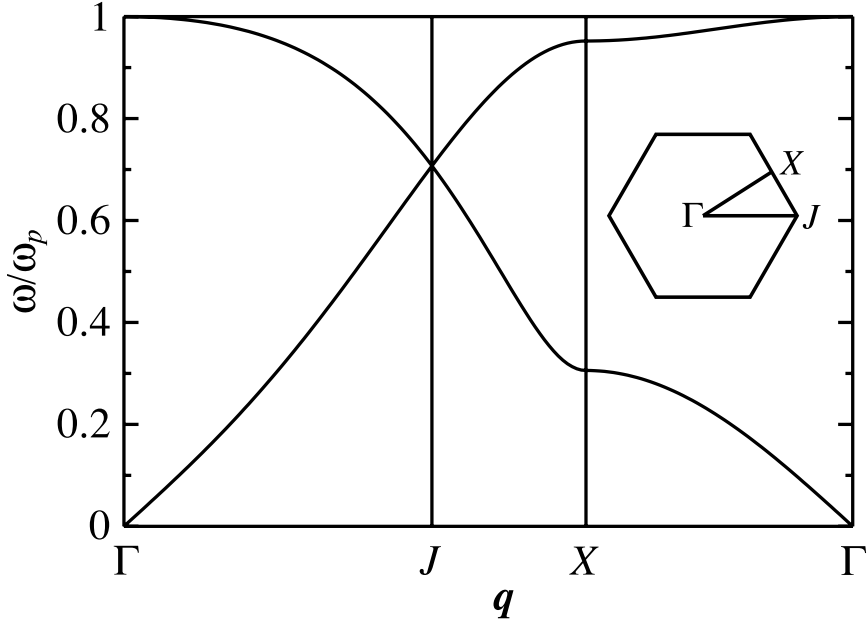}
\caption{Dispersion of the `phonon' modes of the vortex
lattice (with $\omega_c=0$).} \label{figdispersion0}
\end{figure}

\begin{figure}[t]
\centering
\includegraphics[width=4.5in]{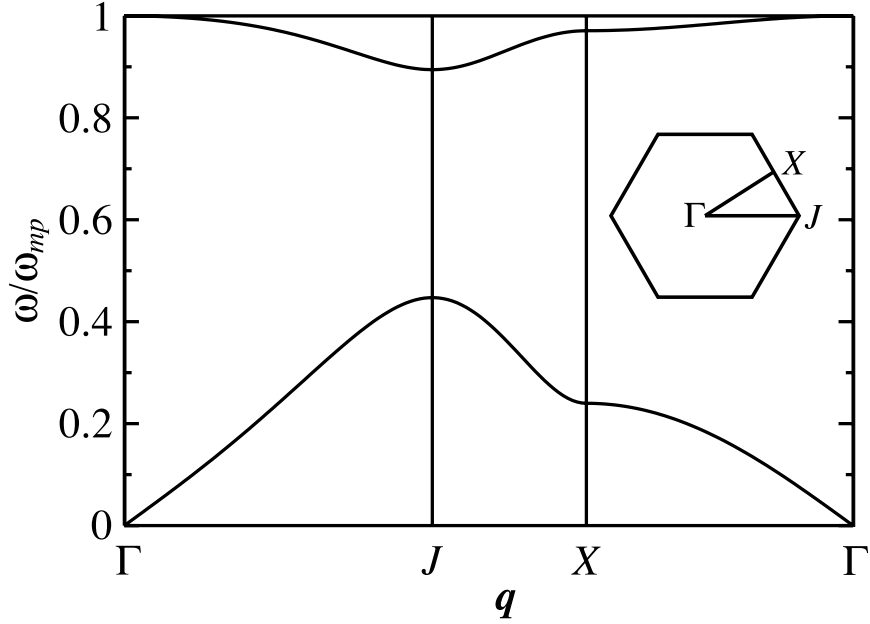}
\caption{Dispersion of the `magnetophonon' modes of the vortex
lattice (here we take $\omega_c=0.5 \omega_p$).} \label{figdispersion}
\end{figure}

We evaluated the dynamical matrix for a perfect triangular lattice
of vortices at positions ${\bf R}_j$ using the Ewald summation
technique (see Appendix \ref{appendix:Ewald}). This leads to the vortex
`phonon' modes shown in Fig.~\ref{figdispersion0} and vortex
`magnetophonon' modes shown in Fig.~\ref{figdispersion}. The
computation of these modes is a generalization of other vortex
oscillation modes discussed previously in superconductors
\cite{degennes,fetter}, rotating superfluids
\cite{fetter1,tkachenko,baym}, and, in a dual picture of `charges',
also to oscillations of electronic Wigner crystals in a magnetic
field \cite{fukuyama,Bonsall77}.

Quantizing these modes, we determine the mean square displacement of
each vortex due to the quantum zero point motion of the vortex
lattice, which we denote $u_{\text{rms}}^2 = \langle |{\bf r}_j -
{\bf R}_j |^2 \rangle/2$.
In terms of the normal modes we find
\begin{align}
        u_{\text{rms}}^2 = \, & \frac{\hbar}{2m_v \omega_p} \nonumber \\
        & {} \times
        \frac{\omega_p A_0}{2} \int_{1\text{BZ}} \frac{d^2q}{(2\pi)^2} \,
        \left[ \frac{1}{\omega_{1}({\bf q})}
        +  \frac{1}{\omega_{2}({\bf q})}
        - \frac{\omega_c^2}{\omega_{1}({\bf q})\omega_{2}({\bf q}) [ \omega_{1}({\bf q}) + \omega_{2}({\bf q}) ]} \right]\;.
\label{eq:standarddeviation}
\end{align}
Here, the momentum integral is over the first Brillouin zone.
The prefactor $\hbar/2m_v\omega_p$ should be identified as the mean square deviation of the position of  a simple one-dimensional oscillator of mass $m_v$ and frequency $\omega_p$ from its equilibrium position.

\begin{figure}[t]
\centering
\includegraphics[width=4.5in]{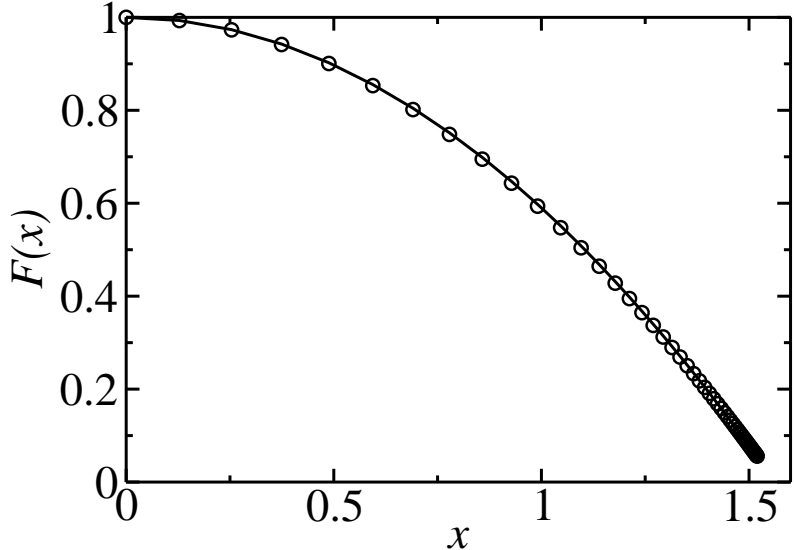}
\caption{Plot of our numerical data (open circles) and fit to our interpolation
formula $F(x)$ (straight line) as given by Eq.~(\ref{e7}).}
\label{fig:F}
\end{figure}

We found an excellent fit (see Fig.~\ref{fig:F})
of our numerical data to the interpolation formula
\begin{equation}
m_v = \frac{0.03627 a_v^2 \hbar^2}{\rho_s u_{\text{rms}}^4} F(x),
\label{e6}
\end{equation}
where $a_v$ is the separation between nearest neighbor vortices,
$x \equiv |\alpha| d u_{\text{rms}}^2/\hbar$, and
\begin{equation}
  F(x) \approx 1 - 0.4099 x^2 \;. \label{e7}
\end{equation}
The above fit is motivated by a simple analytic calculation in which the two modes $\omega_{1,2}({\bf q})$ are replaced by their long-wavelength approximations (see Appendix \ref{appendix:Debye} for details).
Eq.~(\ref{e7}) holds only as long as the r.h.s. is positive, while
$F(x)=0$ for larger $x$ (we will see below and in
Fig.~\ref{graphF} that this apparent upper bound on $x$ is relaxed
once we allow for viscous damping). Similarly, for the frequency
$\omega_{mp}$ in Fig.~\ref{figdispersion}, we obtain $\omega_{mp}
= \sqrt{\omega_p^2 + \omega_c^2}$ with
\begin{equation}
\omega_p = \frac{35.45 \rho_s u_{\text{rms}}^2}{\hbar a_v^2
[F(x)]^{1/2}}~~;~~\omega_c = \frac{27.57 \rho_s u_{\text{rms}}^2 x
F(x)}{\hbar a_v^2}. \label{e8}
\end{equation}
For the experiments of Ref.~\cite{hoffman} we estimate $\rho_s =
12$ meV \cite{weber}, $u_{\text{rms}} = 20$ \AA\ \cite{urms}, $a =
3.83$ \AA, and $a_v=240$ \AA. The overall scale for $m_v$ is
determined by setting $n_S=n_{MI}$ so that $x=0$ and $F(x)=1$.
This yields $m_v \approx 8 m_e$ and $\hbar \omega_p \approx 3$ meV
(or $\nu_p \approx 0.7$ THz). For a more accurate determination,
we need $n_S$, for which there is considerable uncertainty {\em
e.g.\/} for $|n_S - n_{MI}| = 0.015$, we find $x=1.29$, $m_v
\approx 3 m_e$ and $\hbar \omega_p \approx 5$~meV.

\section{Limitations}

We now consider the influence of a variety of effects which have
been neglected in the computation of Section~\ref{sec:vlat}. The
extensions were already discussed in Section~\ref{sec:ft}, and here
we will make quantitative estimates.

\subsection{Viscous drag}
\label{sec:viscous}

\begin{figure}
\centering
\includegraphics[width=4.5in]{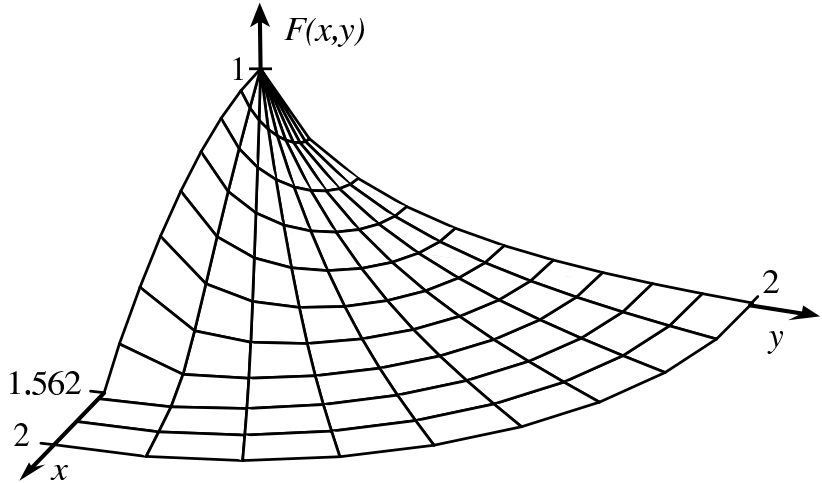}
\caption{Plot of the function $F(x,y)$ which replaces $F(x)$ in
Eqs. (\ref{e6},\ref{e7},\ref{e8}) upon including viscous drag,
$\eta$ ($y \equiv \eta d u_{\rm rms}^2 /\hbar$). The argument $x$
measures the Hall drag $\alpha$ ($x \equiv |\alpha| d u_{\rm
rms}^2 /\hbar$) and the Hall angle is determined by $|\tan
\theta_H | = x/y$.} \label{graphF}
\end{figure}

It is conventional in models of vortex dynamics at low frequencies
\cite{bardeen} to include a dissipative viscous drag term in the
equations of motion, contributing an additional force
\begin{equation}
{\bf F}_D (j) = - \eta d \frac{d {\bf r}_j}{d t}, \label{e9}
\end{equation}
to the r.h.s. of Eq.~(\ref{e2}). This leads to the transformation
Eq.~(\ref{Deta}) in the dynamical matrix. There are no reliable
theoretical estimates for the viscous drag co-efficient, $\eta$, for
the cuprates. However, we can obtain estimates of its value from
measurements of the Hall angle, $\theta_H$, which is given by
\cite{bardeen,ong2} $\tan \theta_H = \alpha/\eta$. Harris {\em et
al.} \cite{ong2} observed a dramatic increase in the value $|\tan
\theta_H|$, to the value 0.85, at low $T$ in ``60 K''
YBa$_2$Cu$_3$O$_{6+y}$ crystals, suggesting a small $\eta$, and weak
dissipation in vortex motion. For our purposes, we need the value of
$\eta$ for frequencies of order $\omega_p$, and not just in the d.c.
limit. The very na\"ive expectation that $\eta(\omega)$ behaves like
the quasiparticle microwave conductivity would suggest it decreases
rapidly beyond a few tens of GHz, well below $\omega_p$
\cite{turner03}. Lacking solid information, we will be satisfied
with an estimate of the influence of viscous drag obtained by
neglecting the frequency dependence of $\eta$ (a probable
overestimate of its influence). The resulting corrections to
Eqs.~(\ref{e6}-\ref{e8}) are easily obtained (as in
Ref.~\cite{blatter}), and can be represented by the replacement
\begin{equation}
F(x) \rightarrow F(x,y),~\mbox{where}~y \equiv \frac{\eta d
u_{\text{rms}}^2}{\hbar}
\end{equation}
and
\begin{equation}
F(x,0)=F(x).
\end{equation}
The sketch of the function $F(x,y)$ is in Fig.~\ref{graphF}; as long
as $y>0$, we have $F(x,y)>0$.
As expected, the viscous damping decreases the estimate of the mass,
and this decrease is exponential for large $y$, {\em e.g.\/} at
$x=0$ we have the interpolation formula
\begin{equation}
F(0,y)
\approx (1 + 0.41 y + 2.69y^2) e^{-3.43 y}.
\end{equation}

\subsection{Meissner screening}

The interaction in Eq.~(\ref{e3}) is screened at long distances by
the supercurrents, and the intervortex coupling becomes
exponentially small. This does have an important influence at small
momenta in that the shear mode of the vortex lattice disperses as
\cite{degennes,fetter} $\sim q^2$. However, as long as $a_v \ll
\lambda$, there will not be a significant influence on
$u_{\text{rms}}$ or $\omega_{p}$.

\subsection{Retardation}

The interaction Eq.~(\ref{e3}) is assumed to be instantaneous; in
reality it is retarded by the propagation of the charged plasmon
mode of the superfluid, and these effects were included in
Eqs.~(\ref{u2}) and (\ref{dabd}). We can estimate the corrections
due to this mode in a model of superfluid layers coupled by the
long-range Coulomb interaction. In physical terms, we compare the
energy per unit area of a `phase fluctuation' at the wavevector of
the vortex lattice Brillouin zone boundary ($\sim \rho_s /a_v^2$)
with its electrostatic energy ($\sim \hbar^2 \omega_p^2 /(e^2 d)$);
this shows that such corrections are of relative order $\sim
(\hbar^2/m_v)/(e^2 d)$. Alternatively this ratio can be viewed as
the order of magnitude of the two terms in the denominator of the
fourth term in Eq.~(\ref{dabd}). For the parameters above and
$d=7.5$ \AA\ this ratio is $\sim 0.009$, and hence quite small.

\subsection{Nodal quasiparticles}

We expect that nodal quasiparticles contribute to the viscous drag,
and so their contribution was already included in the experimentally
determined estimate of $\eta$ in ({\em i\/}). The nodal
quasiparticle contribution to $m_v$ and $\eta$ has recently been
discussed at some length in Ref.~\cite{predrag}. This analysis finds
an infrared finite correction to $m_v$, and a contribution to $\eta$
which vanishes as $T \rightarrow 0$. The latter observations are
consistent with the observations of Harris {\em et al.} \cite{ong2}.

\subsection{Disorder}

We have assumed here a triangular lattice of vortices. In reality,
STM experiments show significant deviations from such a structure,
presumably because of an appreciable random pinning potential. This
pinning potential will also modify the vortex oscillation
frequencies and its mean square displacement. Both pinning and
damping $\eta$ tend to reduce vortex motion.  For this reason, the
estimates of $m_v$ above in which these effects are neglected must
be regarded as upper bounds.

The above considerations make it clear that new experiments on
cleaner underdoped samples, along with a determination of the
spatial dependence of the hole density (to specify $\alpha$), are
necessary to obtain a more precise value for $m_v$; determining
the $H$ dependence of $m_v$ will enable confrontation with theory.

\section{Implications}

An important consequence of our theory is the emergence of
$\omega_p$ as a characteristic frequency of the vortex dynamics. It
would therefore be valuable to have an inelastic scattering probe
which can explore energy transfer on the scale of $\hbar \omega_p$,
and with momentum transfer on the scale of $h/a_v$, possibly by
neutron \cite{neutron} or X-ray scattering; observation of a
resonance at such wavevectors and frequencies, along with its
magnetic field dependence, could provide a direct signal of the
quantum zero-point motion of the vortices. A direct theoretical
consideration of magnetoconductivity in our picture would have
implications for far-infrared or THz spectroscopy, allowing
comparison to existing experiments \cite{Lihn96}; further such
experiments on more underdoped samples would also be of interest.

Another possibility is that the zero point motion of the vortices
emerges in the spectrum of the LDOS measured by STM at an energy
of order $\hbar \omega_p$. We speculate that understanding the
`vortex core states' observed in STM studies \cite{fischer,pan}
%of Bi$_2$Sr$_2$CaCu$_2$O$_{8+\delta}$ and YBa$_2$Cu$_3$O$_{7-\delta}$
will require accounting for the quantum zero point motion of the
vortices; it is intriguing that the measured energy of these
states is quite close to our estimates of $\hbar \omega_p$.

\section{Acknowledgements}

We thank J.~Brewer, E.~Demler, \O.~Fischer, M.~P.~A.~Fisher,
W.~Hardy, B.~Keimer, N.~P.~Ong, T.~Senthil, G.~Sawatzky, Z.~Te\v
sanovi\'c and especially J.~E.~Hoffman and J.~C.~Seamus Davis for
useful discussions. This research was supported by the NSF under
grants DMR-0457440 (L.~Balents), DMR-0098226 and DMR-0455678
(S.S.), the Packard Foundation (L.~Balents), the Deutsche
Forschungsgemeinschaft under grant BA 2263/1-1 (L.~Bartosch), and
the John Simon Guggenheim Memorial Foundation (S.S.).

\appendix

\section{Evaluation of `magnetophonon' modes}
\label{appendix:Ewald}

In this appendix we show how the `phonon' and `magnetophonon' modes
as depicted in Figs.~\ref{figdispersion0} and \ref{figdispersion} can be calculated
using the well-known Ewald summation technique (see e.g. Ref.~\cite{Ziman}).
Our calculation is similar to existing calculations and we will make contact
with earlier work by considering a generalized potential energy of the form
\begin{equation}
  U = \frac{g^2}{2} \sum_{i \neq k} \frac{1}{p | {\bf r}_i - {\bf r}_k |^p} \;,
\end{equation}
with $p>0$.
Here, ${\bf r}_i$ is the position of the $i$'th `charge' (which could also be a real charge)
and denoting ${\bf R}_i$ as is its equilibrium position we write ${\bf r}_i = {\bf R}_i + {\bf u}_i$.
To minimize the total potential energy the ${\bf R}_i$ form a triangular Bravais lattice.
For $p=1$, $U$ reduces to the potential energy of a two-dimensional Wigner crystal
with charges interacting via the three-dimensional Coulomb interaction.
We are particularly interested in the case $p \to 0$ where the interaction between
`charges' becomes logarithmic. This case applies to our vortex lattice:
Taking the gradient of $U$ with respect to ${\bf r_j}$, we obtain the
`electric' force ${\bf F}_E(j) = - {\bf \nabla}_{{\bf r}_j} U$ on the $j$'th
vortex from
\begin{equation}
{\bf F}_E (j) = g^2 \sum_{k (\neq j)} \frac{{\bf r}_j - {\bf
r}_k}{|{\bf r}_j - {\bf r}_k |^{p+2}}
\end{equation}
after setting $p = 0$. Identifying $g^2 = 2 \pi \rho_s$, this equation clearly
reduces to Eq.~(\ref{e3}).

By considering arbitrary $p$ we will now generalize a calculation by Bonsall and Maradudin
\cite{Bonsall77}.
First, we expand $U$ in the displacements from the equilibrium positions $u_{i\alpha}$
(with $\alpha = x,y$ labeling the two cartesian coordinates) and
only keep terms up to second order,
\begin{equation}
  U  =  U_0 + \frac{m_v}{2} \sum_{i\alpha j\beta} \Omega^2_{i\alpha;j\beta}\, u_{i\alpha} u_{j\beta} \;.
  \label{eq:harmonicapproximation}
\end{equation}
To determine the normal modes we essentially have to diagonalize the matrix
\begin{equation}
\Omega^2_{i\alpha;j\beta} \equiv \frac{1}{m_v} \left(\frac{\partial^2}{\partial u_{i\alpha} u_{j\beta}} U \right)_{{\bf u} = 0} \;.
\end{equation}
Fourier transforming $\Omega^2_{i\alpha;j\beta}$ to momentum space gives us a block-diagonal
 matrix $\Omega^2_{\alpha\beta}({\bf q})$ where each block is a $2 \times 2$ matrix.
The action for the vortices is then given by Eq.~(\ref{u1}) with the dynamical matrix
\begin{equation}
  \label{eq:DMApp}
  D_{\alpha\beta}({\bf q},\omega) = A_0 m_v \omega^2 \delta_{\alpha\beta}
  + A_0 B \omega \epsilon_{\alpha\beta} + A_0 m_v \Omega_{\alpha\beta}^2({\bf q}) \;.
\end{equation}
 Due to the long-range interaction all matrix elements $\Omega_{\alpha\beta}^2({\bf q})$ are slowly converging sums which we evaluate using the Ewald summation technique \cite{Ziman}.

 Let us write $\Omega^2_{\alpha\beta}({\bf q})$ as
 \begin{equation}
  \Omega^2_{\alpha\beta}({\bf q}) = - \frac{g^2}{m_v} \left[ S_{\alpha\beta}({\bf q}) - S_{\alpha\beta}(0) \right] \;,
\end{equation}
where the matrix elements $S_{\alpha\beta}({\bf q})$ are defined as
\begin{equation}
  S_{\alpha\beta}({\bf q}) = \frac{\partial^2}{\partial x_{\alpha} \partial x_{\beta}} \left.
  \sum_{{\bf R}_j \neq 0} \frac{e^{-i {\bf q} \cdot {\bf R}_j}}{p | {\bf x} - {\bf R}_j |^p} \right|_{{\bf x} = 0} \;.
\end{equation}
We can now use the integral representation
\begin{equation}
  \frac{1}{p y^p} = \frac{\epsilon^{p/2}}{2 \Gamma(1+p/2)} \int_0^{\infty} dt\, t^{p/2-1}\, e^{-y^2 \epsilon t} \;,
\end{equation}
(with arbitrary $\epsilon > 0$) and divide the integral on the r.h.s.\ in one part with $0 < t < 1$ and one part with $t > 1$.
Setting $y = | {\bf x} - {\bf R}_j|$ we see that the sum over the ${\bf R}_j$ in the integral from $1$ to infinity converges rapidly.
The usual trick is to use Ewald's generalized theta function transformation
\begin{equation}
  \sum_{{\bf R}_j}  e^{-i {\bf q} \cdot {\bf R}_j - |{\bf x} - {\bf R}_j|^2 \epsilon t}
  = \frac{\pi}{A_0 \epsilon t} \sum_{\bf G} e^{i ({\bf G} + {\bf q}) \cdot {\bf x} - {|{\bf G} + {\bf q}|^2}/{4 \epsilon t}} \;,
\end{equation}
%(where the sum over ${\bf G}$ is a sum over all vectors of the reciprocal lattice)
and transform the integrand of the integral from $0$ to $1$ to a sum over the reciprocal lattice ${\bf G}$ such that this sum also converges rapidly.
We then obtain
\begin{align}
S_{\alpha\beta}({\bf q}) = \, & \frac{\epsilon^{p/2}}{2 \Gamma(1+p/2)}  \biggl[
   \sum_{{\bf R}_j} e^{-i {\bf q} \cdot {\bf R}_j}
   \left( 4 \epsilon^2 R_{j\alpha} R_{j\beta} \varphi_{1+p/2} ({\bf R}_j^2 \epsilon)
   - 2 \epsilon \delta_{\alpha,\beta} \varphi_{p/2} ({\bf R}_j^2 \epsilon) \right)
   \nonumber \\
  & \ + \frac{4  \epsilon \delta_{\alpha,\beta}}{2+p}
  - \frac{\pi}{A_0 \epsilon} \sum_{\bf G} (q_{\alpha} + G_{\alpha})(q_{\beta} + G_{\beta})
  \varphi_{-p/2}(|{\bf q} + {\bf G}|^2/4\epsilon) \biggr] \;,
\end{align}
where
\begin{equation}
  \varphi_{\nu}(z) = \int_1^{\infty} dt\, t^{\nu} e^{-zt}
\end{equation}
is a Misra function. The matrix $\Omega_{\alpha \beta}^2({\bf q})$ is now given by
\begin{align}
  \Omega_{\alpha \beta}^2({\bf q}) =\, & \frac{g^2 \pi \epsilon^{p/2-1}}{2 \Gamma(1+p/2) m_v A_0}
 \biggl[ \sum_{\bf G} (q_{\alpha} + G_{\alpha})(q_{\beta} + G_{\beta})
  \varphi_{-p/2}(|{\bf q} + {\bf G}|^2/4\epsilon) \nonumber \\
 & \qquad \qquad -  \sum_{{\bf G} \neq 0}  G_{\alpha} G_{\beta} \, \varphi_{-p/2}({\bf G}^2/4\epsilon) \biggr] \nonumber \\
 {}  + \, & \frac{g^2 \epsilon^{p/2}}{2 \Gamma(1+p/2) m_v} \sum_{{\bf R}_j} \left[ 1 -\cos({\bf q} \cdot {\bf R}_j) \right]
   [ 4 \epsilon^2 R_{j\alpha} R_{j\beta} \, \varphi_{1+p/2} ({\bf R}_j^2 \epsilon) \nonumber \\
 & \qquad \qquad {} - 2 \epsilon \delta_{\alpha,\beta} \varphi_{p/2} ({\bf R}_j^2 \epsilon) ] \;.
 \label{eq:Omegaij}
\end{align}
Setting $p=1$ we recover Eq.~(3.10) of Ref.~\cite{Bonsall77} as we should expect. The case $p \to 0$ is of interest to us. For our vortex lattice we have $g^2 = 2\pi \rho_s$. Using
\begin{align}
  \varphi_0(z) = \, & \frac{1}{z} \, e^{-z} \;, \\
  \varphi_1(z) = \, & \left( 1 + \frac{1}{z} \right) \frac{1}{z} \, e^{-z} \;,
\end{align}
we see that when letting $\epsilon \to 0$ our dynamical matrix evaluated here agrees
with the dynamical matrix given in Eq.~(\ref{u2}) if we neglect retardation effects.
We can now go ahead and calculate the `phonon' or `magnetophonon' dispersion using
\begin{equation}
        \omega_{1,2}^2 =  \frac{(\Omega_{11}^2+\Omega_{22}^2+\omega_c^2)}{2}
     \mp \sqrt{\frac{({\Omega_{11}^2+\Omega_{22}^2+\omega_c^2})^2}{4} - \Omega_{11}^2 \Omega_{22}^2 + \Omega_{12}^2 \Omega_{21}^2} \;.
        \label{eq:omega1and2}
\end{equation}
Here $\omega_c = B/m_v = \pi (n_s - n_{MI})  / a^2 m_v$ is the `cyclotron' frequency.
Identifying the plasma frequency
\begin{equation}
  \omega_p = \left(\frac{2 \pi g^2}{m_v A_0} \right)^{1/2}
\end{equation}
as the characteristic frequency we can evaluate $\omega_1({\bf q})$ and $\omega_2({\bf q})$ using Eq.~(\ref{eq:Omegaij}).
It turns out that the sums over ${\bf R}_j$ and ${\bf G}$ indeed converge rapidly and that $\omega_{1,2}({\bf q})$ are indeed independent of the value of $\epsilon$.
A plot of the spectrum for $n_s = n_{MI}$ (zero `magnetic' field $B$) is shown in Fig.~\ref{figdispersion0}.
Turning on the `magnetic' field $B$ leads to an avoided crossing of the two modes. This can be seen in Fig.~\ref{figdispersion}  where we have chosen $\omega_c = 0.5 \, \omega_p$.

Finally we would like to note that in the long-wave length limit the spectrum becomes isotopic and we find (with $a_v$ being the distance between nearest neighbor vortices)
\begin{align}
  \omega_1 ({\bf q}) & \sim \frac{3^{1/4}}{\sqrt{32 \pi}} \, \frac{\omega_p^2}{\omega_{mp}}  \, (a_v q) \;,
  \label{eq:omega1} \\
  \omega_2 ({\bf q}) & \sim \omega_{mp}  \;,
  \label{eq:omega2}
\end{align}
where $\omega_{mp} = \sqrt{\omega_p^2 + \omega_c^2}$ is the `magnetophonon' frequency. With or without the `magnetic' field the shear mode is always linear in $q$.

\section{Simple Debye model and beyond}
\label{appendix:Debye}

It is instructive to evaluate $u_{\text{rms}}^2$
in the Debye approximation where we replace $\omega_1({\bf q})$ and $\omega_2({\bf q})$ by Eqs.~(\ref{eq:omega1}) and (\ref{eq:omega2}). Also, as usual we replace the first Brillouin zone by a Debye sphere of the same volume.
Using Eq.~(\ref{eq:standarddeviation}) we obtain in this approximation
\begin{equation}
        u_{\text{rms}}^2 = (5/2) \, \hbar / 2m_v\omega_{mp} \;,
        \label{eq:u2Debye}
\end{equation}
which for $B=0$ simplifies to
$u_{\text{rms}}^2 = (5/2) \, \hbar / 2m_v\omega_p$.
To this the shear mode $\omega_1({\bf q})$ contributes 80\%.
Solving for the inertial mass of the vortex we now obtain
\begin{equation}
        m_v = \frac{25 \sqrt{3} \, \hbar^2 a_0^2}{128 \pi^2 \rho_s u_{\text{rms}}^4}
        - \frac{\sqrt{3} a_0^2 B^2}{4 \pi^2 \rho_s} \;.
        \label{eq:MassVortexDebye}
\end{equation}
More accurately, we can evaluate the integral in Eq.~(\ref{eq:standarddeviation}) for the exact dispersion relation numerically.
Extracting the zero field result we have
\begin{equation}
        u_{\text{rms}}^2 = 2.5718 \cdot \frac{\hbar}{2m_v\omega_p} \, I (B/m_v \omega_p)\;,
\label{eq:urms}
\end{equation}
with $I(0)=1$. The numerically determined prefactor $2.5718$ corresponds to the factor $5/2$ in the Debye approximation. The small deviation is mainly due to the fact that as can be seen in Fig.~\ref{figdispersion0} both $\omega_1({\bf q})$ and $\omega_2({\bf q})$ are on average slightly overestimated by Eqs.~(\ref{eq:omega1}) and (\ref{eq:omega2}).
Solving for the mass of a vortex we find
\begin{equation}
        m_v = \frac{(2.5718)^2 \sqrt{3} \, \hbar^2 a_0^2}{32 \pi^2 \rho_s u_{\text{rms}}^4} \,
        F(u_{\text{rms}}^2 B/\hbar) \;.
        \label{eq:massvortex}
\end{equation}
If we define $I_2(z) \equiv z^{1/2} I(z^{1/2})/2$ then the normalized function $F$ is related to the inverse of $I_2$ by $F(x) = x^2/4 I_2^{-1}(x)$ and satisfies $F(0)=1$.
While for the case of the exact dispersion relation considered here $F(x)$ has to be calculated numerically, Eq.~(\ref{eq:MassVortexDebye}) suggests a fit of the form $F(x) =  1 - c_1 x^2$. As can be seen in Fig.~\ref{fig:F} the quality of such a fit turns out to be excellent.


\begin{thebibliography}{99}
\bibitem{urms}  We estimate $u_{\text{rms}}$ from the published
data in Ref.~\cite{hoffman}. Also, using the correspondence between
the LDOS modulations and the vortex core states \cite{fischer}, we
obtain a similar number from the measurements of Ref.~\cite{pan}.
\end{thebibliography}
\end{document}